\documentclass[preprint,12pt]{revtex4-1}

\usepackage{amsmath}    
\usepackage{graphicx}   
\usepackage{verbatim}   
\usepackage{color}      
\usepackage{subfigure}  
\usepackage{hyperref}   
\usepackage{mathrsfs}

\begin{document}

\title{Full phase stabilization of a Yb:fiber femtosecond frequency comb via high-bandwidth transducers}

\author{C. Benko$^{1}$}
\email{craig.benko@jila.colorado.edu}
\author{A. Ruehl$^{2,3}$, M. J. Martin$^{1}$, K. S. E. Eikema$^{3}$, M. E. Fermann$^{2}$, I. Hartl$^{2}$, J. Ye$^{1}$}
\affiliation{
$^{1}$JILA, National Institute of Standards and Technology and University of Colorado, Boulder, Colorado 80309-0440, USA \\
$^{2}$IMRA America, Inc., 1044 Woodridge Avenue, Ann Arbor, Michigan 48105, USA\\
$^{3}$LaserLaB Amsterdam, De Boelelaan 1081, 1081HV Amsterdam, The Netherlands
}

\date{\today}

\begin{abstract}
We present full phase stabilization of an amplified Yb:fiber femtosecond frequency comb using an intra-cavity electro-optic modulator and an acousto-optic modulator.  These transducers provide high servo bandwidths of 580 kHz and 250 kHz for $f_{rep}$ and $f_{ceo}$, producing a robust and low phase noise fiber frequency comb.  The comb was self-referenced with an $f-2f$ interferometer and phase locked to an ultra-stable optical reference used for the JILA Sr optical clock at 698 nm, exhibiting 0.21 rad and 0.47 rad of integrated phase errors (over 1 mHz -- 1 MHz) respectively.  Alternatively, the comb was locked to two optical references at 698 nm and 1064 nm, obtaining 0.43 rad and 0.14 rad of integrated phase errors respectively.  
\end{abstract}

\maketitle{}

\noindent
One of the most important and demanding applications of a femtosecond (fs) frequency comb (FC) has been in high-precision frequency metrology where phase stabilization of carrier envelope offset ($f_{ceo}$) and the repetition rate ($f_{rep}$) are required to analyze and disseminate optical frequency standards~\cite{ludlow2008}.  Fiber-based frequency combs are desirable due to robust and high power operation with low phase noise~\cite{schibli2008,newburry2007}, and ability to connect optical standards to telecom bands in a single step~\cite{ruehl2011}.  Eliminating phase noise in fiber based frequency combs is also essential for coherent high harmonic generation~\cite{cingoz2011a}.

Frequency combs used for optical clockwork are actively stabilized using an optical frequency reference and self-referencing, relying on feedbacks to pump power of the oscillator and piezo-electric transducer (PZT)-actuated cavity mirrors to work on $f_{ceo}$ and $f_{rep}$.  This method of stabilization has proven to be successful for fiber-based FCs and can support mHz relative optical line-widths~\cite{schibli2008} and coherent optical transfer over more than one spectral octave~\cite{ruehl2011}. Recent success in eliminating broadband phase noise in a Yb:fiber oscillator~\cite{cingoz2011} and broad bandwidth of electro-optical modulators (EOMs) in Er:fiber based oscillators~\cite{swann2011,newbury2005,nakajima2011,hudson2005,zhang2010} make it desirable to apply these techniques for a Yb:fiber oscillator.
\begin{figure}[htb]
\centering
\centerline{\includegraphics[width=11cm]{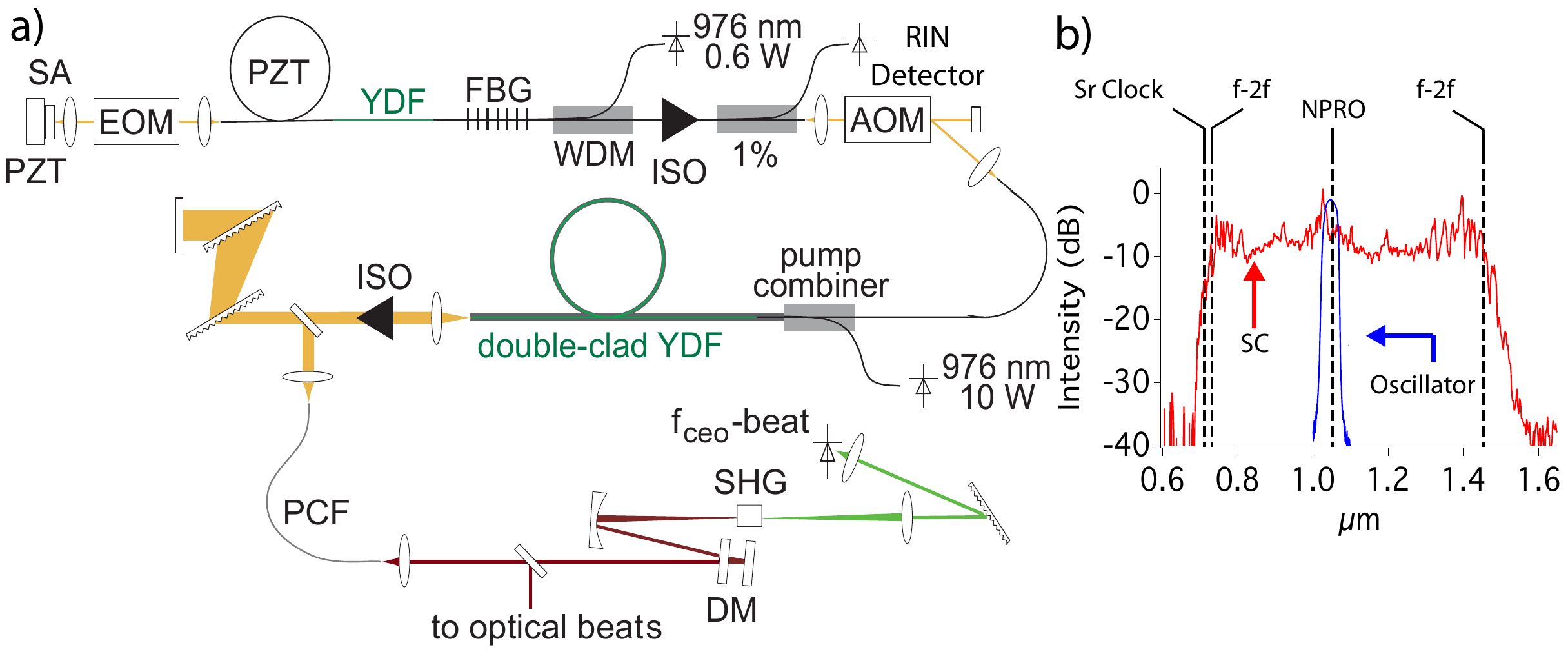}}
\caption{(a) A schematic of the oscillator.  The oscillator includes an intra-cavity EOM and an AOM before the linear chirped pulse amplifier.  (b) The oscillator output spectrum, shown in blue, is centered at 1050 nm and spans $\sim{}$50 nm. A highly nonlinear fiber with flattened dispersion at 1050 nm is used to generate octave spanning spectra, shown in red. YDF, ytterbium doped fiber; SA, saturable absorber; PCF, photonic crystal fiber; SHG, second harmonic generation; DM, dichroic mirror; ISO, optical isolator; FBG, fiber bragg grating; WDM, wavelength division multiplexer; SC, supercontinuum; NRPO, 1064 nm laser.}
\label{fig:fig1}
\end{figure}

In this Letter, we report on full phase stabilization of a Yb:fiber fs FC using an intra-cavity EOM and an extra-cavity acousto-optic modulator (AOM). This combination gives the largest simultaneous servo bandwidth for both $f_{ceo}$ and $f_{rep}$ stabilization, providing operational flexibility and long-term robustness and making the fiber comb valuable for optical clocks. The use of external AOM, which acts only on $f_{ceo}$~\cite{jones2000}, also provides an orthogonalization between the two feedback loops.  The achieved phase noise integrated over a large bandwidth (1 mHz -- 1 MHz) is among the lowest on record.

The FC used in our experiment, sketched in Fig.~\ref{fig:fig1}, is based on a Fabry-Perot-type Yb-similariton oscillator mode-locked with a sub-ps lifetime saturable absorber mirror and dispersion compensated by a chirped fiber Bragg grating for operation in the similariton regime~\cite{hartl2005}. We introduced an anti-reflection coated 1 mm long LiTaO${}_{3}$ crystal to the free-space section of the laser cavity to serve as an EOM. The net laser cavity dispersion was balanced close to zero, but slightly positive to preserve mode-locking in the similariton regime for increased passive stability.

To minimize amplitude-to-phase noise conversion generated by self phase modulation in the fiber amplifier, we used a linear chirped-pulse amplification scheme.  To this end, we obtained 2 W of power and sub-80 fs pulses with  $f_{rep}=$ 168 MHz, determined by frequency-resolved optical gating.  After amplification, the light was launched into a 18 cm piece of highly nonlinear, flattened dispersion fiber centered at 1050 nm (NKT Photonics NL-1050-ZERO-2).  This produced nearly flat, octave-spanning spectra, making it an ideal tool to interface with current optical clocks, including the Sr optical lattice clock operating at 698 nm~\cite{ludlow2008}. Fig.~\ref{fig:fig1} contains a schematic of the oscillator, the oscillator spectrum and the super-continuum.
\begin{figure}[htb]
\centerline{\includegraphics[width=9cm,height=8.5cm]{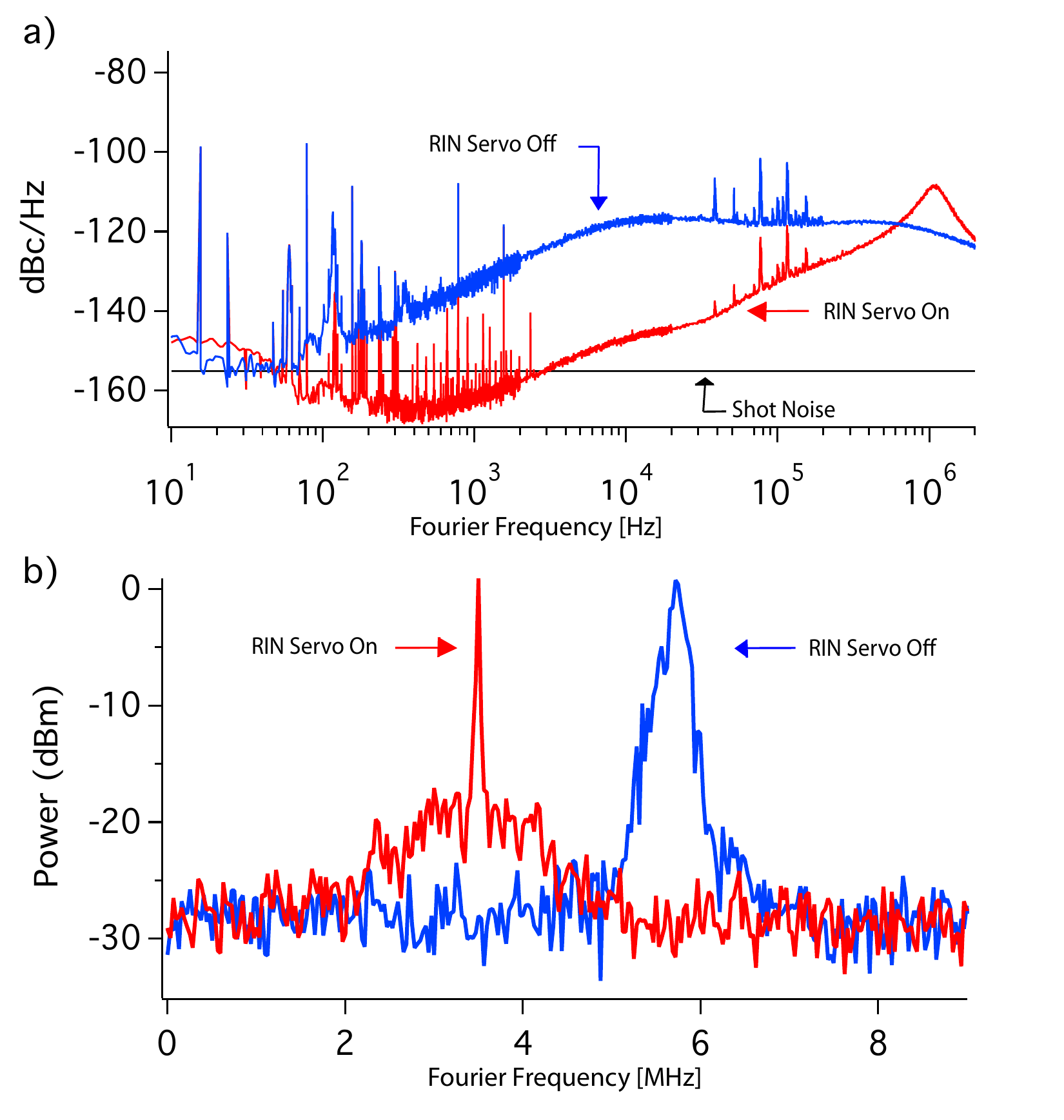}}
\caption{(a) The RIN PSD is shown with the servo on/off.  (b) The free running $f_{ceo}$ linewidth dramatically narrows when the intensity noise is reduced, limited by a 10 kHz resolution bandwidth (RBW).}
\label{fig:fig2}
\end{figure}

Before we fully stabilized the FC, we reduced the residual intensity noise (RIN) of the oscillator.  Similar in approach to Ref.~\cite{cingoz2011}, a home-built fiber-coupled detector with a flat frequency response to 4 MHz was used to detect the intensity noise.  The correction signal was filtered and applied to the current of the oscillator pump diode. By reducing RIN from the oscillator, the free running line-width of the $f_{ceo}$ beatnote was dramatically narrowed from 70 kHz to sub-10 kHz, an effect not previously observed from RIN reduction in the oscillator.  To our knowledge, this is the narrowest and most stable free running $f_{ceo}$ beatnote from a fs FC.  We also observed reduced phase noise on optical beats at 698 and 1064 nm.  Fig.~\ref{fig:fig2} shows the RIN servo performance demonstrating substantial intensity noise reduction with a unity gain crossing at $\sim{}$600 kHz, as well as the free running $f_{ceo}$ signals with the RIN servo on and off, respectively.

To verify the performance of the intra-cavity EOM as a frequency transducer, a single comb tooth was loosely phase locked via PZT to a 1064 nm nonplanar ring oscillator (NPRO) with $f_{ceo}$ free running.  The EOM was driven by a high voltage power supply that can produce 1 kV peak to peak modulation up to 1 MHz.  Modulations outside the servo bandwidth were applied and the power in the modulation sidebands were measured on the same heterodyne beat signal.  From 38 kHz to 500 kHz, the power in the modulation sideband exhibits a $\frac{1}{\omega^2}$ dependence, where $\omega$ is the modulation frequency.  This is consistent with the expected $J_{1}^{2}(\beta)$ dependence, where $\beta$ is the phase modulation index and $J_{1}^{2}(\beta)$ is the first order Bessel function.  Using the electro-optic coefficients of LiTaO$_{3}$ from~\cite{casson2004}, the amplitude of phase modulation per applied voltage agreed to within 10$\%$ of the expected value. Using the RIN detector, it was verified that there was negligible amplitude modulation generated by the EOM. A single comb tooth was successfully phase locked to the 1064 nm NPRO using the EOM, and the PZT was used to keep the EOM within its dynamic range. A servo bandwidth of $\sim{}$390 kHz was obtained, limited by a piezo electric resonance in the EOM at $\sim{}$580 kHz.

Pump current modulation primarily affects $f_{ceo}$ but can also change $f_{rep}$ via various nonlinear mechanism~\cite{newbury2005}. In our system, the pump current was used to reduce intensity noise of the oscillator, making it unavailable for dynamic $f_{ceo}$ control. The addition of an AOM was necessary in order to provide an ideal $f_{ceo}$ actuator.  To obtain phase-lock, the $f_{ceo}$ beat was bandpassed and sent to a digital phase detector to generate an error signal.  The signal was filtered and sent to a voltage-controlled oscillator centered at 200 MHz to drive the AOM.  Slow feedback was applied to the temperature of the fiber Bragg grating, used as the output coupler of the oscillator, to tune the cavity dispersion and thus $f_{ceo}$. A servo bandwidth of 250 kHz was obtained using the AOM, most likely limited by acoustic wave propagation delay in the AOM.

\begin{figure}[htb]
\centerline{\includegraphics[width=9cm,height=9cm]{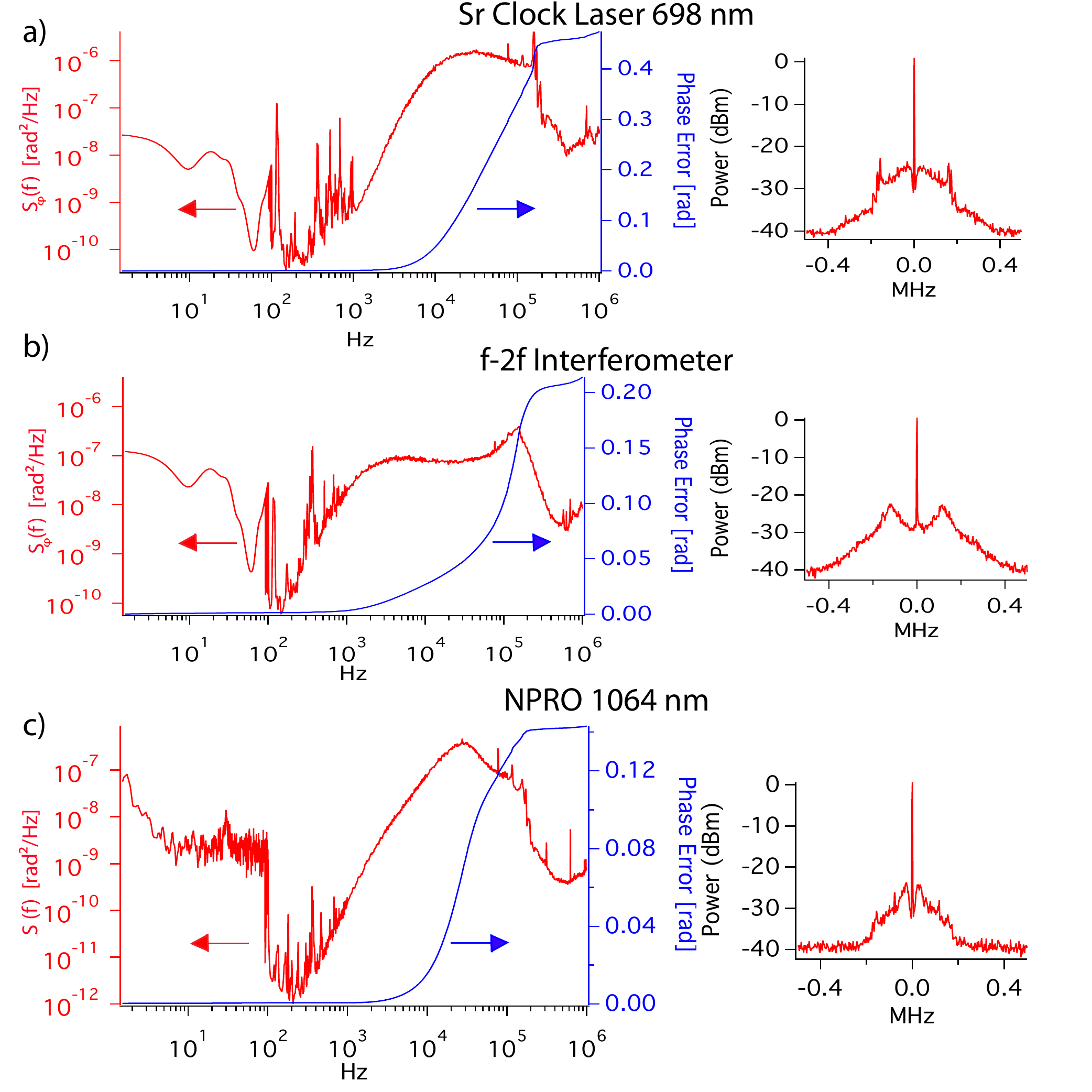}}
\caption{The phase noise PSD (left column, left axis), the integrated RMS phase noise (right axis), and the heterodyne beat spectra (right column) of the in-loop phase-lock errors. In (a) and (b), full phase stabilization is achieved with optical phase-lock at 698 nm (a) and self-referencing phase-lock of $f_{ceo}$ (b).  In (c), full stabilization is achieved with optical phase-locks at 1064 nm and 698 nm, and the 1064 nm in-loop analysis is shown.  The phase noise integration was performed from 1 mHz to 1 MHz.  The beat signals are shown with 1 kHz RBW.}
\label{fig:fig3}
\end{figure}

To fully stabilize the FC, $f_{ceo}$ and $f_{rep}$ need to be controlled simultaneously. $f_{ceo}$ was stabilized with the AOM.  Instead of directly stabilizing $f_{rep}$, a single comb tooth was phase locked to the JILA Sr optical lattice clock laser at 698 nm~\cite{swallows2011} using the intra-cavity EOM and a PZT mirror. The beat signal was bandpassed and a digital phase detector provided the error signal. To analyze the servo performance, the in-loop phase noise power spectral density (PSD), $S_{\phi}(f)$ [rad$^{2}$/Hz], was measured and integrated to determine the total phase error.  The $f_{ceo}$ and 698 nm phase lock contain 0.21 rad and 0.47 rad residual phase error respectively when integrated from  1 mHz to 1 MHz. Fig.~\ref{fig:fig3}~(a,b) displays the phase error PSD, integrated RMS phase noise, and beat spectra of the phase-locks.  Using the relation~\cite{hall1992},
\begin{align}
	\eta = e^{- \int{S_{\phi}(f) df}},
\end{align}
\noindent where $\eta$ is the fractional power remaining in the coherent carrier of the phase-lock, the $f_{ceo}$ phase-lock exhibits $\eta = 0.96$ and the 698 nm phase-lock exhibits $\eta = 0.8$. When integrated from  1 mHz to 1 kHz, the 698 nm and $f_{ceo}$ have phase errors of 1.6 mrad and 3.3 mrad, showing excellent phase noise suppression around the carrier.

To further demonstrate the versatility of the FC, we implemented an equivalent method for stabilization by phase locking two comb teeth to two different optical references, forgoing the use of an $f-2f$ interferometer~\cite{ye2000}.  By locking two independent comb teeth to two optical references, both comb degrees of freedom are stabilized. To test stabilization performance using this method, we phase locked the FC to both the 698 nm ultra-stable laser and a 1064 nm NPRO.  The EOM/PZT actuator phase locked the 698 nm beat and the AOM phase locked the 1064 nm beat.  Fig.~\ref{fig:fig3}~(c) contains the phase noise PSD and the beat spectrum of the 1064 nm phase-lock. The 698 nm lock exhibited similar performance when co-stabilized with $f_{ceo}$.  Again, low integrated phase noise was achieved with 0.43 rad and 0.14 rad of phase error in the 698 nm and 1064 nm phase-locks respectively, when integrated from 1 mHz to 1 MHz.  This corresponds to both locks having $\eta=0.83 \text{ and } 0.98$. This scheme demonstrates that the FC transfers coherence throughout the entire spectrum, making it well suited for interfacing with various optical standards.

In conclusion, we present a fully phase stabilized Yb:fiber FC using an intra-cavity EOM and extra-cavity AOM.  We used current modulation of the oscillator pump diode to reduce RIN while still maintaining dynamic control of $f_{ceo}$ via the AOM.  An intra-cavity EOM was successfully introduced inside the laser cavity, providing an extra high bandwidth actuator on optical phase-locks.  Two equivalent methods of complete phase stabilization were evaluated with excellent locking performance, demonstrating that optical coherence can be established anywhere within the generated supercontinuum, and thus making it a versatile tool for frequency metrology. Funding for this work is provided by NIST and DARPA. A. Ruehl acknowledges support from a European Community Marie Curie Fellowship.


\begin{thebibliography}{99}
	\bibitem{ludlow2008} A. D. Ludlow et al., Science 319, 1805-1808 (2008).
	\bibitem{schibli2008}T. R. Schibli, I. Hartl, D. C. Yost, M. J. Martin, A. Marcinkevicius, M. E. Fermann, and J. Ye, Nature Photonics 2, 355-359 (2008).
	\bibitem{newburry2007}N. R. Newbury and W. C. Swann, JOSA B 24, 1756–1770 (2007).
	\bibitem{ruehl2011}A. Ruehl, M. Martin, K. Cossel, L. Chen, H. McKay, B. Thomas, C. Benko, L. Dong, J. Dudley, M. Fermann, I. Hartl, and J. Ye, Phys. Rev. A 84, 1050-2947 (2011).
	\bibitem{cingoz2011a}A. Cingoz, D. C. Yost, T. K. Allison, A. Ruehl, M. E. Fermann, I. Hartl, and J. Ye.  Nature (in press) (2012).
	\bibitem{cingoz2011}A. Cingoz, D. C. Yost, T. K. Allison, A. Ruehl, I. Hartl, M. E. Fermann, and J. Ye, Opt. Lett. 36, 743-745 (2011).
	\bibitem{hudson2005}D. D. Hudson, K. W. Holman, R. J. Jones, S. T. Cundiff, J. Ye, and D. J. Jones, Opt. Lett. 30, 2948-2950 (2005).
	\bibitem{newbury2005}N. Newbury and B. R. Washburn, IEEE J. Quant. Elec. 41, 1388-1402 (2005).
	\bibitem{nakajima2011}Y. Nakajima, H. Inaba, K. Hosaka, K. Minoshima, A. Onae, M. Yasuda, T. Kohno, S. Kawato, T. Kobayashi, and T. Katsuyama,  Optics Express 18, 1667–1676 (2010).
	\bibitem{zhang2010}W. Zhang, Z. Xu, M. Lours, R. Boudot, Y. Kersalé, G. Santarelli and Y. Le Coq, Appl. Phys. Lett., 96, 211105 (2010).
	\bibitem{swann2011}W. Swann, E. Baumann, and F. Giorgetta, Optics Express 19, 24387-24395(2011).
	\bibitem{jones2000}R. J. Jones and J. C. Diels,  Phys. Rev. Lett. 86, 3288-3291 (2001).
	\bibitem{hartl2005} I. Hartl, G. Imeshev, L. Dong, G. C. Cho, and M. E. Fermann,
	in CLEO/QELS Technical Digest (Optical Society of America, 2005), paper CThG1. 
	\bibitem{swallows2011}M. Swallows,M. J. Martin, M. Bishof, C. Benko, Y. Lin, S. Blatt, A. M. Rey and J. Ye, IEEE EFTF/TUFFC, TUFFC-04585-2011.R1 (2011)
	\bibitem{casson2004}J. L. Casson, K. T. Gahagan, D. A. Scrymgeour, R. K. Jain, J. M. Robinson, V. Gopalan, and R. K. Sander, JOSA B 21, 1948-1952 (2004).
	\bibitem{hall1992}J. L. Hall and M. Zhu, International School of Physics ‘Enrico Fermi’, Course CXVIII, Laser Manipulation of Atoms and Ions (E. Arimondo, W.D. Phillips, and F. Strumia, Eds., North Holland, 1992), pp. 671-702.
	\bibitem{ye2000}J. Ye, J. L. Hall, and S. A. Diddams, Opt. Lett. 25, 1675-1677 (2000).
\end{thebibliography}
\end{document}